\documentclass[a4paper,11pt]{article}
\pdfoutput=1

\newif\ifjournal
\journalfalse

\usepackage{longtable}
\usepackage{booktabs}
\usepackage{jheppub}
\usepackage[T1]{fontenc}

\makeatletter
\ifjournal\else\gdef\@fpheader{}\fi
\def\LT@makecaption#1#2#3{%
  \LT@mcol\LT@cols c{\hbox to\z@{\hss\parbox[t]\LTcapwidth{%
    \reset@font\small
    \sbox\@tempboxa{#1{#2. }#3}%
    \ifdim\wd\@tempboxa>\hsize
      #1{#2. }#3%
    \else
      \hbox to\hsize{\hfil\box\@tempboxa\hfil}%
    \fi
    \endgraf\vskip\baselineskip}%
  \hss}}}
\makeatother

\allowdisplaybreaks[1]
\hypersetup{pdftitle={The next-to-next-to-leading order BFKL eigenvalue at odd
  conformal spin in planar N=4 super Yang-Mills},
  pdfauthor={A. Prygarin and C. C. Madjuogang Sandeu},
  pdfkeywords={BFKL, conformal spin, harmonic sums, N=4 super Yang-Mills,
  Regge trajectory}}

\newcommand{\zb}{\bar{z}}                       

\newcommand{\cpl}{a}                            
\newcommand{\chiord}[1]{\chi_{#1}}              
\newcommand{\Fn}{F_{n}}                          

\newcommand{\Sh}[2]{S_{#1}\!\left(#2\right)}         
\newcommand{\Sig}[1]{\Sigma_{#1}}               
\newcommand{\Del}[1]{\Delta_{#1}}               

\newcommand{\Aset}{\mathcal{A}}                 
\newcommand{\atom}[1]{A_{#1}}                   
\newcommand{\Ratom}{\mathcal{R}}               
\newcommand{\BAatom}{\mathcal{B}}              
\newcommand{\coeff}[1]{c_{#1}}                 

\newcommand{\lntwo}{\ln 2}
\newcommand{\zzt}{\zeta_2\zeta_3}

\newcommand{\ap}{a_{+}}
\newcommand{\am}{a_{-}}
\newcommand{\amax}{a_{\max}}
\newcommand{\amin}{a_{\min}}
\newcommand{\Gladder}{G}
\newcommand{\epsk}{(-1)^{k}}                     
\newcommand{\wgt}[1]{\mathcal{W}_{#1}}           
\newcommand{\Kres}{\mathcal{K}}
\newcommand{\Lam}{\Lambda}
\newcommand{\Lax}[1]{\Lambda^{\max}_{#1}}
\newcommand{\Lin}[1]{\Lambda^{\min}_{#1}}         
\newcommand{\Lnul}[1]{\Lambda^{0}_{#1}}           
\newcommand{\Psiw}{\Psi}

\newcommand{\KRone}{K_{\mathcal{R}1}}
\newcommand{\KRtwo}{K_{\mathcal{R}2}}
\newcommand{\KRthree}{K_{\mathcal{R}3}}
\newcommand{\Om}[1]{\Omega_{#1}}               
\newcommand{\cOm}{c_{O}}                       
\newcommand{\cCm}{c_{C}}                       
\newcommand{\dAB}{c_{AB}}                      

\newcommand{\Nshapes}{forty-one}
\newcommand{\Nshapesrat}{five}
\newcommand{\Nshapesnest}{thirty-six}
\newcommand{\Nclosing}{forty}
\newcommand{\Ncomposed}{thirty-seven}
\newcommand{\Nfamilies}{forty-seven}
\newcommand{\Nresidual}{seven}
\newcommand{\Nweights}{seven}
\newcommand{\Ncomposites}{five}
\newcommand{\Ncompterms}{forty-five}
\newcommand{\Nomega}{four}

\title{\boldmath The next-to-next-to-leading order BFKL eigenvalue at odd conformal
spin in planar \texorpdfstring{$\mathcal{N}=4$}{N=4} super Yang--Mills}

\author[a,1]{A.~Prygarin\note{Corresponding author.}}
\author[a]{and C.~C.~Madjuogang Sandeu}

\affiliation[a]{Department of Physics, Ariel University, Ariel 40700, Israel}

\emailAdd{alexanderp@ariel.ac.il}

\abstract{
We give the next-to-next-to-leading order color-singlet BFKL eigenvalue of planar
$\mathcal{N}=4$ super Yang--Mills at odd conformal spin $n$ in closed form. At
the two lower orders the known expressions contain nested harmonic sums only at
the two conjugate points $z=(|n|-1)/2+i\nu$ and $\zb$. At three loops the sums
are evaluated on a ladder of integer-shifted arguments from the reflected
point $-\zb$ up to $z$, with one point past it, the coefficient at a rung fixed
by its two distances to the ends. Seven families are the exception, their
coefficients written through ladder sums whose summand shifts with the summation
index. For one of the seven the coefficient lies outside the fixed-coefficient algebra
of nested harmonic sums of the two distances, an arithmetic obstruction in the
denominators. That exclusion reaches the eigenvalue coefficient only through an
agreement of rule and coefficient verified and not proved. Rules uniform in the
conformal spin produce all forty-seven
families at every odd $n\ge3$; no per-spin coefficient is tabulated in the
closed form, and the $n=1$ boundary block is supplied
separately. Those rules are a reconstruction from the computed spins, exact at every one
of them, verified over the range $n\le99$ and at the holdout spin $n=101$, and not
proved at arbitrary odd $n$. Each atom-table coefficient is a rational combination of $1$,
$\pi^{2}$ and $\zeta_3$, coefficient and atom together carrying transcendental weight
five. Along $\nu=0$ the intercepts match the Quantum Spectral Curve values at each
computed odd spin through $n=91$, the extent of those values. The comparison tests the
integrand and the reduction together at one point of each spin, for most of them for the
first time. Away from that line the closed
form gives the collinear behavior of the block uniformly in the spin, which the
intercepts alone do not fix.
}

\keywords{BFKL, conformal spin, harmonic sums,
$\mathcal{N}=4$ super Yang--Mills, Regge trajectory}

\arxivnumber{2607.17608}

\begin{document}
\maketitle

\section{Introduction}
The high-energy limit of planar $\mathcal{N}=4$ super Yang--Mills is controlled by the
BFKL kernel, whose eigenvalue on the principal series of the M\"obius group is a
function of the conformal spin $n$ and a variable $\nu$, where the combination
$\gamma=\tfrac12+i\nu$ is the anomalous-dimension
variable~\cite{Fadin:1975cb,Kuraev:1977fs,Balitsky:1978ic,Lipatov:1985uk}. Its
multi-Reggeon extension is a completely integrable system in multicolor QCD, in the
generalized leading-logarithmic
approximation~\cite{Lipatov:1993yb,Faddeev:1994zg}. At leading and next-to-leading
order the eigenvalue is a combination of nested harmonic sums evaluated at the two
conjugate points $z=(|n|-1)/2+i\nu$ and $\zb=(|n|-1)/2-i\nu$, the leading eigenvalue
being $\chiord{0}=-[\Sh{1}{z}+\Sh{1}{\zb}]$~\cite{Fadin:1975cb,Kuraev:1977fs,%
Balitsky:1978ic,Lipatov:1985uk,Kotikov:2002ab}. That form is given at $n=0$ in
Ref.~\cite{Costa:2012cb} and at any $n$ in App.~C of Ref.~\cite{Alfimov:2018cms}. The next-to-leading
eigenvalue~\cite{Fadin:1998py,Ciafaloni:1998gs} obeys the maximal-transcendentality
principle of Kotikov and Lipatov, which fixes it, for the dimensional-reduction (DRED)
coupling used here, as the highest-weight part of the corresponding QCD
result~\cite{Kotikov:2002ab,Kotikov:2000pm,Kotikov:2006ts}. It is
already hermitian separable in their generalized sense, additively separable apart from
one mixed
term in the two conjugate points~\cite{Kotikov:2002ab,Bondarenko:2015tba,Joubat:2020hvc}.
That mixed term is a property of the way Kotikov and Lipatov write the eigenvalue,
since their equation for the corresponding Bethe--Salpeter kernel is additively
separable at any spin, the eigenvalue entering that equation implicitly.

The three-loop kernel was obtained by Caron-Huot and
Herranen~\cite{Caron-Huot:2016tzz}. What they give is the linearized kernel in
coordinate space in closed form, a radial integrand obtained symbolically at one
integer spin at a time, harmonic-sum expressions at $n=0$ and at $n=1$, and
numerical trajectories above those; a compact reduced formula in harmonic sums, holding
uniformly in the conformal spin away from the line $\nu=0$, is absent from that
work. Those authors state that for even conformal spin $n=2,4,6,\dots$ the radial
integrand needs a generalization of harmonic polylogarithms and the resulting
eigenvalue is not a combination of conventional harmonic
sums. A prediction for the non-conformal piece of the planar QCD kernel at that order
has since been given~\cite{Brunello:2025rhh}. Along the line $\nu=0$ much more is
known. At $n=0$ the eigenvalue is in closed form from the Quantum
Spectral Curve~\cite{Gromov:2013pga,Alfimov:2014bwa,Gromov:2015vua}, with an independent
reconstruction proposed shortly after~\cite{Velizhanin:2015xsa}. Velizhanin
later reconstructed the pomeron eigenvalue one logarithmic order
higher~\cite{Velizhanin:2021bdh}. Away from $n=0$,
Alfimov, Gromov and Sizov gave the intercepts in closed form for arbitrary conformal
spin, in terms of binomial harmonic sums of uniform transcendental weight
five~\cite{Alfimov:2018cms}. The Regge trajectories have since been treated by an
asymptotic Baxter--Bethe method~\cite{Ekhammar:2024neh,Ekhammar:2025vig}. Off the line
$\nu=0$ the harmonic-sum expressions stop at those two
spins~\cite{Caron-Huot:2016tzz,Gromov:2015vua}. A Bethe--Salpeter ansatz fixed the terms of highest depth at arbitrary
$\nu$ and $n$~\cite{Joubat:2020vrw,Joubat:2020hvc}, leaving the simpler ones open and,
as noted there, differing at $n=1$ from Ref.~\cite{Caron-Huot:2016tzz}; a formula
uniform in the spin for the eigenvalue itself is still lacking. That difference is not
settled here: the three-loop input is taken from Ref.~\cite{Caron-Huot:2016tzz}
throughout, the $n=1$ block included. The adjoint, color-octet eigenvalue at
this order is known in closed form uniform in the conformal spin, up to constants left
undetermined there~\cite{Dixon:2012yy} and fixed later from the multi-Regge limit of
the four-loop remainder function~\cite{Dixon:2014voa}. The same eigenvalue was obtained
at finite coupling by continuation from the collinear limit~\cite{Basso:2014pla}, the
eigenvalues of its next-to-leading kernel computed directly~\cite{Fadin:2011we}. For
the color-singlet eigenvalue a closed form of this kind at odd conformal spin has, to
the best of our knowledge, not appeared before.

Information on the line does not determine the eigenvalue off it. At $\nu=0$
the two conjugate points coincide, $z=\zb=(n-1)/2$, the block equals its conjugate,
and the intercept $\chiord{2}(n,0)$ is one half of their common value; at $n\ge3$ that
value is reached through a removable singularity, since $(n-1)/2$ lies at an internal
rung of the ladder of Eq.~\eqref{eq:distances}, where the separate contributions have
poles canceling among themselves inside the block, while at $n=1$ no contribution has
a pole at that point. Once the two points agree, a term in $z$ is a term in $\zb$, so
product-freeness, the absence in the fixed atom class of any term coupling the two,
becomes empty, and the pole structure in $\nu$ is not seen at a fixed $\nu$. The
intercepts therefore fix one point of each trajectory. We give the eigenvalue away from
that point, at odd $n$.

A closed form states the transcendental weight, the pole structure in $\nu$, and
the expansion of the block into the fixed atom class of Eq.~\eqref{eq:atoms},
from which product-freeness follows. Odd $n$ is the
natural first sector, since the next-to-leading anomalous term that violates the
generalized holomorphic separability of Kotikov and Lipatov~\cite{Kotikov:2002ab}, a
property distinct from the hermitian separability above, comes with a prefactor
$1+(-1)^{n}$ and vanishes at odd spin, where the eigenvalue
coincides with its continuation under $|n|\to-|n|$. Those authors note in the same place
that Bose symmetry of the two-gluon wave function leaves even $n$ physical for the
pomeron itself. The $\nu\to0$ limit at odd spin was examined one order lower, where the
eigenvalue reduces to rational multiples of $1$, $\pi^{2}$
and $\zeta_3$~\cite{Joubat:2021nss}. Reflection identities
among harmonic sums connect the eigenvalue with its value at the reflected argument,
and a sequence of studies made that structure
visible~\cite{Prygarin:2018tng,Prygarin:2018cog,Joubat:2019esj,Joubat:2020vrw,%
Joubat:2023ftd}. The ladder found here is
the three-loop answer to the same question. The present study states the closed form, gives in
Appendix~\ref{app:catalog} the rules that produce its coefficients, and reports the checks
those rules pass. The reduction of the three-loop integrand to the atom tables, and the
per-spin record behind the wider comparison, are in Ref.~\cite{companion}. The rules in
machine-readable form accompany this paper.

\section{Definitions}
\label{sec:def}
The Regge trajectory reads
\begin{equation}
\omega(n,\nu)=4\cpl\big[\chiord{0}+\cpl\,\chiord{1}+\cpl^{2}\,\chiord{2}+\cdots\big],
\label{eq:omega}
\end{equation}
with coupling $\cpl=g^{2}=g_{YM}^{2}N_c/(16\pi^{2})$, so that $\chiord{2}$ is the
three-loop eigenvalue. Each order is even in $\nu$, and
the present order is built from a holomorphic block $\Fn$ evaluated at the two points
\begin{equation}
z=\tfrac{|n|-1}{2}+i\nu,\qquad \zb=\tfrac{|n|-1}{2}-i\nu,
\qquad
\chiord{2}=\tfrac14\big[\Fn(z)+\Fn(\zb)\big].
\label{eq:zdef}
\end{equation}
For real $\nu$ the two points are complex conjugates and each order is real. The
nested harmonic sums follow the non-strict
convention~\cite{Vermaseren:1998uu,Gromov:2015vua}, with negative letters alternating.
In the shifted argument $z+k$ at which the atoms are evaluated they are continued off
the integers on the even branch, the continuation from even integer argument used in
Ref.~\cite{Gromov:2015vua}, which makes them meromorphic with poles only at the
negative integers. Each nested atom of Eq.~\eqref{eq:atoms} is fixed by its own sum,
whose nested harmonic sums sit at the integer summation index, so the branch enters
only where an alternating sum at a shifted argument is written in closed form, as in
Eq.~\eqref{eq:merge}. At the two ladder distances of Eq.~\eqref{eq:distances}, which
are non-negative integers at every rung, the nested harmonic sums are the finite
sums. We write
$\Sig{a}=\Sh{a}{z}+\Sh{a}{\zb}$ and $\Del{a}=\Sh{a}{z}-\Sh{a}{\zb}$. An empty
summation range gives zero, $\Sh{a_1,\dots,a_d}{0}=0$ for every non-empty word while
the empty word gives $1$; the rules below are evaluated at zero argument at the two
ends of the ladder, so this value is part of the formula. Here $\nu$ is half the
internal variable of Refs.~\cite{Caron-Huot:2016tzz,Gromov:2015vua}. The eigenvalue
depends on $|n|$ only, and from here on $n$ stands for the positive odd spin.

A rational atom
$1/w^{q}$ has transcendental weight $q$. For a nested atom of Eq.~\eqref{eq:atoms} the weight
collects the weight of the nesting word $v$, the powers $p$ of the summation variable
in its extra denominators, and the pole order $q$. A coefficient inherits the weight
of the constant multiplying it, zero for $1$, two for $\pi^{2}$ and three for
$\zeta_3$, the rational prefactor adding no weight, plus the weight of the harmonic
sums of the two ladder distances dressing it; for a ladder sum of Eq.~\eqref{eq:laddersum} the count
is $j$, plus the weight of whatever harmonic sum appears under the summation.

\section{The closed form}
For every odd $n$ the holomorphic block is a fixed prefactor plus a sum of functions of
one variable,
\begin{equation}
\Fn(z)=-88\,\zeta_4\,\Sh{1}{z}-16\,\zzt-80\,\zeta_5
       +\sum_{a\in\Aset_n}\coeff{a}\,\atom{a}(z),
\label{eq:master}
\end{equation}
with $\Aset_n$ the atom table at spin $n$.

The three-loop input supplies this
prefactor at every spin. Equation~(5.12) of Ref.~\cite{Caron-Huot:2016tzz}
writes the holomorphic block as a Mellin integral plus the two terms
$\gamma_K^{(3)}F^{(1)}-C^{(3)}$, and they stand in front of the sum
above: $\gamma_K^{(3)}=\tfrac{11\pi^{4}}{45}=22\,\zeta_4$ from the cusp anomalous
dimension of that work, $F^{(1)}=-4\,\Sh{1}{z}$ from its one-loop eigenvalue, and
$C^{(3)}=16(\zzt+5\,\zeta_5)$. No explicit $n$ appears in any of the three.

Two kinds of function occur, which we call
atoms, and we write both as functions of their argument $w$,
\begin{align}
(\Ratom)&:\quad \atom{}^{\Ratom}(w)=\frac{1}{w^{q}},\qquad q=1,\dots,5,\nonumber\\
(\BAatom)&:\quad \atom{}^{\BAatom}(w)=\sum_{N\ge1}
   \frac{\sigma^{N}\,\Sh{v}{N}}{\bigl[\prod_{(d,p)}(N+d)^{p}\bigr]\,(N+w)^{q}},
\label{eq:atoms}
\end{align}
with $\sigma=\pm1$, with $\Sh{v}{N}$ a nested harmonic sum, and with $(d,p)$ a short
list of shifts and powers, the shifts non-negative integers, fixed by the shape and zero in every shape that
occurs. There are
\Nshapes{} such functions (\Nshapesrat{} rational, \Nshapesnest{} nested). We call one of them a
shape, and one shape placed at one rung an instance. Paired with the constants
$1$, $\pi^{2}$ and $\zeta_3$, the shapes give \Nfamilies{} coefficient families, one for
each shape and constant that occur together.
A family is called a slot in the ancillary file \texttt{chi\_nnlo\_rules.py}.

The spin enters only through the length of the ladder, which fixes the points at
which the atoms are evaluated and the coefficients that multiply them. Those
points are the shifted arguments $w=z+k$ with $k=-(n-1),\dots,0$, joined by one
further point at $k=1$; an individual atom need not reach every one of them. The
shifts place the arguments on the segment from $-\zb=z-(n-1)$ up to $z$. That
segment is the ladder, each point $z+k$ on it a rung; the point $z+1$ lies one step
outside, and the two distances between a rung and the two ends,
\begin{equation}
\ap=k+n-1,\qquad \am=-k,\qquad \ap+\am=n-1,
\label{eq:distances}
\end{equation}
are the natural variables for the coefficients. Neither the
\Nshapes{} shapes nor the \Nfamilies{} families grow with the spin; growth occurs only
in the count of shifted instances, which is $39n-3$ at
each odd $n\ge3$ in the computed range.

At leading and
next-to-leading order the conjugate points are the only arguments that occur, with a
single mixed bilinear $2\,\Del{1}\Del{-2}$ coupling them. At three loops the arguments
spread over the whole ladder, and the coefficient at each rung is a function of the two
distances of Eq.~\eqref{eq:distances}. For \Nclosing{} of the \Nfamilies{} families that
function is a combination of nested harmonic sums, and we call those \Nclosing{} the
closing families. For the remaining \Nresidual{} we write that function instead through
sums taken along the ladder, and for one of those \Nresidual{} no such combination exists
within the fixed-coefficient algebra of those sums, as Sec.~\ref{sec:coeff} shows.

\section{The coefficients}
\label{sec:coeff}
Every one of the \Nfamilies{} families follows from a rule uniform in the conformal
spin, so that no tabulated coefficient enters the closed form of the eigenvalue at any
odd $n\ge3$. The ancillary files supply the rules as
\texttt{chi\_nnlo\_rules.py}. The rules were fixed at low spin, by reconstruction from
previously computed spins, and they are unproved at arbitrary odd $n$;
Sec.~\ref{sec:checks} gives the range over which they are verified. At the
boundary spin $n=1$ the coefficients are read from the supplied boundary
block, the $n=1$ function of Ref.~\cite{Caron-Huot:2016tzz}, with no reassembly.

When the sum over the atoms is taken before the sum along the ladder, \Ncomposed{} of
the \Nclosing{} merge into \Ncomposites{} functions of one variable under distinct ladder weights, each
weight fixed by the two distances and multiplying its function at every rung,
the three left over standing apart as the rational ladder term and the pair at $k=1$,
\begin{align}
\sum_{a\in\Aset_n}\coeff{a}\,\atom{a}(z)
&=\underbrace{32\,\zeta_5-32\,\Sh{5}{z}}_{\text{the two single-rung families, merged}}
\nonumber\\[4pt]
&\quad+\sum_{k=-(n-1)}^{0}(-1)^{k}\,\Gladder_{(-1)^k}(z+k)
\nonumber\\[4pt]
&\quad+\sum_{k=-(n-1)}^{0}(-1)^{k}\,\Sh{1}{\amax}\,\Phi(z+k)
\nonumber\\[4pt]
&\quad+\sum_{k=-(n-1)}^{0}(-1)^{k}
\Bigl[\Sh{1}{\ap}\,D_{+}+\Sh{1}{\am}\,D_{-}
\nonumber\\[4pt]
&\qquad\qquad\qquad\qquad\quad+\Sh{1}{\amin}\,D_{\min}\Bigr](z+k)
\nonumber\\[4pt]
&\quad+\frac{16\pi^{2}}{3}\!\!\sum_{k=-(n-1)}^{0}\!(-1)^{k}
\frac{\Sh{2}{\ap}+\Sh{-2}{\ap}}{z+k}+\Kres(n,z),
\label{eq:ladderform}
\end{align}
with $\amax=\max(\ap,\am)$ and $\amin=\min(\ap,\am)$, so that
$\Sh{1}{\amax}=\Sh{1}{\ap}+\Sh{1}{\am}-\Sh{1}{\amin}$ and the coefficients stay
inside the same set of functions of the two distances. The \Ncomposites{} functions $\Gladder_{(-1)^k}$,
$\Phi$, $D_{+}$, $D_{-}$ and $D_{\min}$ do not change with the spin, and
$\Gladder_{(-1)^k}$ is one and the same function at every rung, its subscript the
ladder weight under which it stands. Each is a
combination of the atoms of Eq.~\eqref{eq:atoms} at a single argument, written out term
by term in Eq.~\eqref{eq:composites}. $D_{+}$,
$D_{-}$ and $D_{\min}$ come with empty nesting word, reducing to polygammas of halved
argument, together with three rational terms in $D_{+}$, so the nested content of the
\Ncomposed{} rests entirely in
$\Gladder_{(-1)^k}$ and $\Phi$, and in that sense it collapses into two composite
functions. Equation~\eqref{eq:ladderform} retains all \Ncomposites{}, with the rational ladder
term, the merged pair and $\Kres$. The last term
$\Kres(n,z)$ holds the \Nresidual{} families written through sums along the ladder.
Equations~\eqref{eq:zdef}, \eqref{eq:master}, \eqref{eq:ladderform}
and~\eqref{eq:kres}, with the coefficient rules collected in
Appendix~\ref{app:catalog}, are the closed form of the eigenvalue at odd conformal
spin and the main result of this paper.

The residual term reads
\begin{equation}
\begin{aligned}
\Kres(n,z)&=\sum_{k=-(n-2)}^{0}\Bigl[\frac{\KRone(n,k)}{z+k}
   +\frac{\KRtwo(n,k)}{(z+k)^{2}}+\frac{\KRthree(n,k)}{(z+k)^{3}}\Bigr]\\
&\quad+\sum_{k=-(n-1)}^{0}\Bigl[\cOm(n,k)\,\Om{3}(z+k)+\cCm(n,k)\,\Om{C}(z+k)\\
&\qquad\qquad\quad+\dAB(n,k)\bigl(\Om{A}(z+k)+\Om{B}(z+k)\bigr)\Bigr],
\end{aligned}
\label{eq:kres}
\end{equation}
with six coefficient functions acting on seven atoms. The rational atoms stop at
$k=-(n-2)$ and the nested ones go one rung further, to $k=-(n-1)$. These six are the
coefficients at the constant $1$, and whatever $\pi^{2}$ and $\zeta_3$ parts the same
atoms carry come with the other terms of Eq.~\eqref{eq:ladderform}, so a coefficient
function of $\Kres$ is the whole coefficient of its atom only for $\Om{3}$ and for
$1/(z+k)^{3}$. The
\Nomega{} nested atoms are
\begin{equation}
\begin{aligned}
\Om{3}(x)&=\sum_{N\ge1}\frac{1}{N(N+x)}=\frac{\Sh{1}{x}}{x},\\[2pt]
\Om{C}(x)&=\sum_{N\ge1}\frac{1}{N^{2}(N+x)}
 =\frac{\pi^{2}}{6x}-\frac{\Sh{1}{x}}{x^{2}},\\[2pt]
\Om{A}(x)&=\sum_{N\ge1}\frac{(-1)^{N}}{N(N+x)^{2}},\\[2pt]
\Om{B}(x)&=\sum_{N\ge1}\frac{(-1)^{N}}{N^{2}(N+x)} .
\end{aligned}
\label{eq:omegas}
\end{equation}
The last two share one coefficient rule, $\dAB$, and their atoms appear solely in the
combination
\begin{equation}
\Om{A}(x)+\Om{B}(x)=\frac{\Sh{-2}{x}}{x}=\frac1x\Bigl[-\frac{\pi^{2}}{12}
-\frac{\zeta\bigl(2,\tfrac{x+2}{2}\bigr)-\zeta\bigl(2,\tfrac{x+1}{2}\bigr)}{4}\Bigr],
\label{eq:merge}
\end{equation}
the alternating counterpart of $\Om{3}$ in Eq.~\eqref{eq:omegas}, one weight higher.
On the even branch $\Sh{-2}{x}$ is that difference of polygammas of halved argument,
the $\lntwo$ and the shifted-argument Nielsen beta function canceling between the two
atoms. The cancellation stops short of $\Om{C}$, and folding $\Om{C}$ into the same
combination, as $\Om{A}+\Om{B}-\Om{C}$, leaves $\cCm+\dAB$ on $\Om{C}$, free of the
ladder sums of Eqs.~\eqref{eq:cd} and~\eqref{eq:cC}. That regrouping takes no atom out
of the block, and each of the seven atoms keeps a coefficient in front of it, so the
count is over those atoms and not over the independent directions their coefficients
span. There are therefore \Nresidual{} rules,
one to a family, and six of them are distinct. Two of the six,
those on the alternating pair and on $\Om{C}$, read
\begin{align}
\dAB&=(-1)^{k}\Bigl[-32\bigl(\Lam(\ap,\am)+\Lam(\am,\ap)\bigr)-32\,\Sh{1}{\amin}^{2}
\nonumber\\
&\hspace{3.0em}+96\,\Sh{1}{\ap}\Sh{1}{\am}+32\bigl(\Sh{2}{\ap}+\Sh{2}{\am}\bigr)\Bigr],
\label{eq:cd}\\
\cCm&=(-1)^{k}\Bigl[32\bigl(\Lam(\ap,\am)+\Lam(\am,\ap)\bigr)\nonumber\\
&\hspace{3.0em}+192\,\Sh{1}{\amin}\Sh{1}{\amax}
 -32\,\Sh{1}{\amin}^{2}\nonumber\\
&\hspace{3.0em}-32\,\Sh{1}{\amax}^{2}-96\,\Sh{1}{\ap}\Sh{1}{\am}\nonumber\\
&\hspace{3.0em}+64\bigl(\Sh{2}{\ap}+\Sh{2}{\am}\bigr)\Bigr],
\label{eq:cC}
\end{align}
with $\Lam$ the ladder sum of Eq.~\eqref{eq:laddersum} below. A rule is called
exchange-even or exchange-odd according to the sign it takes under
$\ap\leftrightarrow\am$, the reflection of the ladder in its midpoint. The
exchange-odd rule $\cOm$ needs weight three and is given in Eq.~\eqref{eq:cO}; no fit of
their own enters the three rational families, which
follow from Eq.~\eqref{eq:laurent} once the nested coefficients are known.

The two terms at $k=1$ merge into a single Hurwitz zeta,
$32\,\zeta(5,z+2)+32/(z+1)^{5}=32\,\zeta(5,z+1)$, and contribute
$16\,\zeta_5-8\,\Sig{5}$ to the eigenvalue. The block contains no other $\Sh{5}{}$
term. Equation~\eqref{eq:atoms} shows that an $\Sh{5}{}$ arises only from a
pole of order five; the atoms of that order are all found at $k=1$, at each spin
computed.

Three of the remaining \Nresidual{} multiply the rational atoms $1/(z+k)^{q}$, and for one of
those three the failure to close is forced, for a reason that is arithmetic. At
integer argument $m$ a nested harmonic sum of weight $\ell$ has denominator
dividing $\mathrm{lcm}(1,\dots,m)^{\ell}$. Take a polynomial expression in such sums, its arguments
drawn from $\ap$, $\am$ or $\amin$, and its rational coefficients the same at
every spin, so that they share one fixed common denominator. Such an expression
has denominator primes only up to $\amax$, apart from the finitely many
primes of that fixed denominator. Coefficients depending rationally on $n$ or on
the two distances fall outside this hypothesis, and the argument does not exclude
them. The coefficient $\KRthree$ of $1/(z+k)^{3}$ exceeds
that bound. Take $n=p+2$ with $p$ prime and $\am=(p-1)/2$, where $\amax=(p+3)/2$.
The sum of
Eq.~\eqref{eq:KR3} below runs to $N=(p+3)/2$, and only at its last two values of $N$
does the harmonic sum in the final term reach argument $p$ and $p+1$. Those two $1/p$
parts do not cancel, so $p$ divides the denominator of $\KRthree$ at every prime
$p\ge5$. The primes $p$ are unbounded, while the common
denominator of a rule uniform in the spin is fixed once and for all, so no rule of that
kind can supply them. One enlargement, the ladder sums of
Eq.~\eqref{eq:laddersum}, closes the three rational families together, $\KRthree$ by
Eq.~\eqref{eq:KR3} and $\KRone$ and $\KRtwo$ by Eq.~\eqref{eq:laurent}, so the
exclusion of that single family rules out closing the sector in the fixed-coefficient
algebra of sums at the two distances. The
other two are not excluded by it. They do show primes above the bound over nearly the
whole computed range, though without a proof: among the twenty-three spins $n=p+2$
with $p\ge5$ and $p+2\le99$, taken at the same rung $\am=(p-1)/2$, the prime $p$ divides the
denominator of $\KRone$ at all twenty-three and that of $\KRtwo$ at
twenty-one of them, with $p=5$ and $p=47$ the two exceptions at that rung. At both
spins $\KRtwo$ does show $p$ in its denominator at some other rungs, again with
$p>\amax$. The other four families of $\Kres$ come out of
the same reduction, and no prime above the bound appears in any of their coefficients
over the odd spins $21\le n\le99$, so the signature is a property of the family and does
not come from the method. That
is still not a proof for the two rational ones.

The argument bears on the coefficient rule, and only through the rule on the
eigenvalue. Wherever a spin was generated, rule and stored coefficient coincide cell
by cell as rational numbers, so no distinction arises there. Beyond that range the
exclusion concerns the rule; transferring it to the eigenvalue coefficient assumes
that the rule reproduces that coefficient at every odd spin, a property verified
across the whole computed range and not proved. The argument is set out in full in
Ref.~\cite{companion}.

The \Nresidual{} families of $\Kres$ close through sums taken along the ladder, whose summand
contains a harmonic sum evaluated at an argument moving with the summation index,
\begin{equation}
\Lax{j}(\ap,\am)=\sum_{N=1}^{\ap}\frac{\Sh{1}{\max(\am+N,\ap-N)}}{N^{j}},
\qquad \Lam\equiv\Lax{1},
\label{eq:laddersum}
\end{equation}
together with the same sums with $\min$ in place of $\max$. The upper limit is the first
argument, so that $\Lam(\am,\ap)$ runs to $N=\am$. The rules use $j=1$ and
$j=2$ and no more. Most such sums reduce to ordinary nested sums. The part that does
not reduce is a window sum, its two summation limits approaching each other,
\begin{equation}
\Psiw_{j,q}(\ap,\am)=-\sum_{N=1}^{(\amax-\amin)/2}\frac{1}{N^{j}}
   \sum_{i=\amin+N+1}^{\amax-N}\frac{1}{i^{q}},
\qquad \Psiw\equiv\Psiw_{1,1},
\label{eq:window}
\end{equation}
The upper limit of the outer sum is an integer, since $\amax-\amin=|2k+n-1|$ is even
at odd $n$. The window sum vanishes whenever $\amax-\amin\le2$ and satisfies, in the
convention fixed above,
\begin{equation}
\begin{split}
\Lam(\ap,\am)+\Lam(\am,\ap)&=\Sh{1,1}{\ap}+\Sh{1,1}{\am}\\
&\quad+\Sh{1}{\ap}\Sh{1}{\am}-\Psiw(\ap,\am).
\end{split}
\label{eq:collapse}
\end{equation}
The exchange-odd family takes $\Psiw_{2,1}$ in the same way. In these
functions the \Nomega{} families on nested atoms close exactly, and so does the coefficient
of $1/(z+k)^{3}$ at rung $k=-m$, which reads
\begin{equation}
\begin{aligned}
\KRthree(n,-m)&=32\,(-1)^{m}\!\!\sum_{N=1}^{\,n-1-m}\!
\Bigl[\frac{2}{N^{2}}+\frac{(-1)^{N}}{N^{2}}-\frac{\Sh{1}{N}}{N}\\
&\qquad+\frac{\Sh{1}{\max(m+N,\,n-1-m-N)}}{N}\Bigr].
\end{aligned}
\label{eq:KR3}
\end{equation}
The harmonic sum under the summation extends to the full ladder length $n-1$. That
length is the argument the denominator primes require, and the fixed-coefficient
algebra of sums at the two distances cannot reach it. The two remaining
rational families follow from a property of the block itself. At an internal rung,
$z=m$ with $m=0,\dots,n-2$, the
block has no
singularity while its shapes do, each at the rungs of its own
range, and requiring every pole to cancel fixes them. At each such rung
and at each pole order $q$ the coefficient multiplying $(z-m)^{-q}$
vanishes,
\begin{equation}
\begin{aligned}
&\coeff{}^{(\Ratom,q)}(n,-m)
 +\sum_{N=1}^{\,n-1-m}\ \sum_{\text{families of pole order }q}\\
&\hspace{5.5em}
 \coeff{}^{(\BAatom,\sigma,v,\{(d,p)\},q)}(n,-m-N)\,
 \frac{\sigma^{N}\Sh{v}{N}}{\prod_{(d,p)}(N+d)^{p}}\;=\;0 ,
\end{aligned}
\label{eq:laurent}
\end{equation}
and it does so separately for the constants $1$, $\pi^{2}$ and $\zeta_3$. The condition
returns $\KRone$ and $\KRtwo$ at pole
orders one and two once the nested coefficients are known, with $\dAB$ entering at
order one through $\Om{B}$ and at order two through $\Om{A}$, and at
pole order three, where no nested family of $\Kres$ enters, it returns $\KRthree$
outright, Eq.~\eqref{eq:KR3} being its closed form. At the far rung, $m=n-1$ and $k=-(n-1)$, the
condition is empty, since no atom has a pole at $z=n-1$. The point $k=1$ lies outside the
ladder and Eq.~\eqref{eq:laurent} is not asserted there. The fifth-order pole at
$z=-1$ comes from the one shape of that order there, $32/(z+1)^{5}$, which has no
partner to cancel against; at the collinear point $z=-1$ the leading pole of
Eq.~\eqref{eq:collinear} is that same singularity.

\section{Structure}
Every atom-table coefficient is a rational combination of $1$, $\pi^{2}$ and
$\zeta_3$, and no individual family involves a constant of weight above
three. In the grading of Sec.~\ref{sec:def} the weight is split between coefficient
and atom, the coefficient supplying whatever the atom lacks, so neither part on its
own need reach weight five while every product of the two does.

At each fixed odd spin the block expands into the fixed atom class of
Eq.~\eqref{eq:atoms}, whose members are functions of one variable, and is therefore
product-free. Product-freeness is a statement about the class and not
about the decomposition of Eq.~\eqref{eq:zdef}, available at fixed spin for any
function even in $\nu$. The atom poles stop short of the far rung, while
$2\chiord{2}$ taken as the block has a fifth-order pole past it, at $\zb=-1$
($z=n$), so membership in the class is a restriction. Hermitian separability as
Refs.~\cite{Bondarenko:2015tba,Joubat:2020hvc} use the term, with no atom class
entering it, is a different statement, and the present study does not claim it at
this order. Within the class the property is stable, since the operations taking the
atom table to Eq.~\eqref{eq:ladderform} act inside it, while for transformations that
leave the class we claim nothing. The one mixed term of the next-to-leading
eigenvalue, the bilinear $2\,\Del{1}\Del{-2}$, has no counterpart in the atom table.

Near a collinear point $z=-j$, with $j\ge1$, the eigenvalue develops a pole
in the anomalous-dimension variable, and the collinear resummation of the DGLAP
logarithms controls the strength of that pole. The closed form fixes it uniformly
in the spin. With $u=z+j$,
\begin{equation}
\Fn=\frac{32}{u^{5}}
   +\frac{48\,(-1)^{j+1}\bigl[\Sh{1}{n+j-1}-\Sh{1}{j-1}\bigr]}{u^{4}}+O(u^{-3}).
\label{eq:collinear}
\end{equation}
The term $32/u^{5}$ comes from the merged pair at $k=1$, whose Hurwitz zeta
$32\,\zeta(5,z+1)$ has a pole at every collinear point. At $j=0$ the
block is regular, $z=0$ being an internal rung at $n\ge3$, where every pole cancels
by Eq.~\eqref{eq:laurent}, and the far rung at $n=1$, where no atom has a pole. At
$j=1$ Eq.~\eqref{eq:collinear} gives
$\chiord{2}=8/u^{5}+12\,\Sh{1}{n}/u^{4}+O(u^{-3})$. The leading
number is the one predicted by the collinear resummation. A change of the energy scale
shifts the pole by $\omega/2$~\cite{Ciafaloni:1998gs}. Iterating that shifted
pole~\cite{Salam:1998tj} and solving the resulting quadratic gives, in the convention of
Eq.~\eqref{eq:omega}, the coefficients $1$, $-2$, $8$ and $-40$ at successive orders,
of which $8$ is the three-loop one. At $n=0$ the whole polar
part at a given loop order is fixed by the anomalous dimension at that
order~\cite{Marzani:2007gk,Costa:2012cb}, and Eq.~(5.17) of
Ref.~\cite{Caron-Huot:2016tzz} gives that part at three loops.
The intercepts do not fix the spin-dependent $u^{-4}$ coefficient. It follows from the
closed form away from $\nu=0$.

\section{Checks}
\label{sec:checks}
Equations~\eqref{eq:ladderform} and~\eqref{eq:kres}, transcribed term by term into
\texttt{chi\_nnlo\_rules.py} and expanded back into the atom basis, reproduce all
$112112$ coefficients of the per-spin tables in the ancillary files of
Ref.~\cite{companion} at odd $n=3,\dots,99$ and a further $4538$ at $n=101$, every one
of them exact in rational arithmetic, the full set of \Nfamilies{} families covered.
Against the tables supplied here the comparison reaches nineteen of those spins, odd
$n=3,\dots,33$ together with $n=51$, $71$ and $91$, and $22412$ coefficients, all of them
exact in rational arithmetic and three of the nineteen at spins above those the closing
rules were fixed on. Rebuilt from the rules one family at a time,
with no tabulated input at any point and Eq.~\eqref{eq:ladderform} nowhere in the route,
the atom table matches the stored one entry for entry at each of those nineteen spins, so
there the eigenvalue put together in the two ways is a single number at every $\nu$;
agreement to a stated number of digits never enters. The expanded equations are compared
with the stored table in \texttt{check\_compact\_form.py}, and the two routes with each
other, at the spins named above, as \texttt{README.txt} sets out.
Equations~\eqref{eq:ladderform} and~\eqref{eq:kres} taken at $n=1$, where $\ap=\am=0$ at
the rung $k=0$ and only the merged single-rung families and $\Gladder_{(-1)^k}$ survive,
return all twenty-eight coefficients of the boundary block, exact in rational
arithmetic; that block is the harmonic-sum expression of Caron-Huot and
Herranen~\cite{Caron-Huot:2016tzz}, whose three-loop integrand we take
as input throughout.

The two sides of this comparison have no table and no program in common:
the stored tables come out of
the exact Mellin extraction from the three-loop
integrand, an extraction in which no closed form appears anywhere, while the other
side is the printed equation, with no stored table behind it. They do share their origin, since the
rules were inferred from tables at lower spins produced from that same integrand. The
comparison therefore probes the reconstruction and the extrapolation of it in the
spin, and no independent input is tested. For two of the families, though, $\KRone$ and
$\KRtwo$ follow from Eq.~\eqref{eq:laurent} once the nested coefficients are known, so at
those two families the comparison tests that identity on the stored table.

We fixed the rules
for the \Nclosing{} closing families on the spins $n\le49$ and then tested them at
$n=93,95,97,99$,
whose coefficients had not yet been computed. They reproduce all $14580$ of them. The
\Nresidual{} remaining rules were fixed at low spin in the same way and written afterward,
so the coefficients of those \Nresidual{} at those four spins serve as a consistency check
only, without predictive force.

One further spin, the holdout spin $n=101$, was generated after all \Nfamilies{} rules
had been written down. Its atom table was produced by the same extraction from the
three-loop integrand,
with $3936$ shifted instances and $4538$ coefficients over the \Nfamilies{} families. The
rules reproduce each entry exactly, and no rule is left without an entry.
Equation~\eqref{eq:laurent} holds there too, at each of its $1500$ Laurent coefficients
(three constants at five pole orders on each of the $100$ internal rungs), and no rule
enters that part of the check. Eight of the fifteen combinations of constant and pole
order occur nowhere on the ladder, so $800$ of them are absent from both sides. One more
vanishes by cancellation and not by absence, at $\pi^{2}$ and pole order one on the
internal rung $m=n-2$: there $\ap=1$, so the factor $\Sh{2}{\ap}+\Sh{-2}{\ap}$ of the
rational ladder term of Eq.~\eqref{eq:ladderform} is zero, while the nested side reaches
only the $N=1$ term, at the far rung $k=-(n-1)$, where the two $\pi^{2}$ families of
pole order one differ only in $\sigma$ and enter with equal coefficients. Both sides
vanish at $801$ coefficients, and the identity has content at the other $699$. All of it
is one spin, a test of the rules and no proof at arbitrary odd $n$.

Along $\nu=0$ the closed form agrees with the Quantum Spectral Curve values of
Ref.~\cite{Alfimov:2018cms} at every odd spin from $n=3$ to $n=91$, where their data set
ends; Table~\ref{tab:intercepts} gives a sample of $\chiord{2}(n,0)$. The reference side
is read from that data set, as the $g^{6}$ coefficient of $\omega(n,0)$, divided by
four; we have not re-evaluated their binomial-sum formula. At $n=1$ the supplied
boundary block returns the reference value, zero, to the accuracy of the
evaluation. Caron-Huot and Herranen fixed the constant $C^{(3)}$ of
Eq.~\eqref{eq:master} by imposing that the eigenvalue vanish there, so at that spin the
agreement follows from the normalization of the input and does not test the reduction
performed here. It is worth mentioning that the
three-loop kernel of Ref.~\cite{Caron-Huot:2016tzz} had been matched to the Quantum
Spectral Curve at $n=0$, and that Alfimov, Gromov and Sizov report agreement between
that kernel and their own closed form at several of the first non-negative spins as
well~\cite{Alfimov:2018cms}. Neither comparison reaches the reduction performed here, so
the agreement tests the integrand and the reduction as one object, at one point of each
spin. For most of those spins no such comparison had been made before. In the reverse
direction, the closed form of Ref.~\cite{Alfimov:2018cms} came out of adjusting a basis
of binomial harmonic sums, at odd spin, to Quantum Spectral Curve data. That basis was
narrowed by uniform transcendentality and by a reciprocity conjecture, which the same
authors find violated one order higher. The route taken here starts from the three-loop
integrand, a different starting point that supports their closed form.

Off the line $\nu=0$ the closed form is compared with the notebook attached to
Ref.~\cite{Caron-Huot:2016tzz}, which evaluates their three-loop trajectory at integer
spin and any $\nu$. Their spectral variable is twice the one used here, so
$\chiord{2}(n,\nu)$ is one quarter of \texttt{j3Eval} at $n$ and $2\nu$, and at
$(n,\nu)=(3,0.5)$, $(5,0.5)$, $(7,0.5)$, $(9,0.3)$ and $(11,0.7)$ the two numbers agree
to at least twenty-six significant digits. The difference there is set by the working
precision of that notebook at those five points, thirty digits, and it
does not bound the agreement. All five of those spins lie inside the range $n\le49$ on
which the closing rules were fixed, so the same comparison was made at $n=51$, $71$, $91$
and the holdout spin $n=101$, at $\nu=0.3$, $0.5$ and $0.7$ of each. The two numbers
agree to thirty-one significant digits at $n=51$, twenty-three at $n=71$, fifteen at
$n=91$ and eleven at $n=101$, the same count at the three values of $\nu$ of a given
spin.

The notebook sets the limit at every one of those twelve points, and the limit tightens
as the spin grows. At $(n,\nu)=(51,0.3)$ it agrees with the closed form to eight,
fifteen, twenty-three and thirty-one significant digits at thirty, forty,
fifty and sixty digits of working precision, so what the difference measures there is its
own truncation. The agreements above are at sixty. At thirty digits the higher spins are
out of its reach altogether. At $(n,\nu)=(91,0.3)$ it returns about $-1.1\times10^{6}$,
where the value is $-266.53$, and prints that number to full precision with no warning. The
closed form is evaluated at fifty digits, and moving the summation cutoff from $200$
to $300$ leaves at least its first thirty-two digits unchanged at each of the four
spins, more in every case than the notebook holds.

Nothing of the reduction performed here enters the notebook side, so at four spins above
the range the closing rules were fixed on, and at three values of $\nu$ at each of them,
the extrapolation in the spin is tested away from $\nu=0$ and not only at the single
point of each spin that an intercept supplies. No proof at arbitrary odd $n$ follows from
twelve points. Elsewhere off the line the check rests on the coefficient comparison,
which fixes the block at every $\nu$ and not at one point.

\begin{table}[t]
\centering
\begin{tabular}{@{}rll@{}}
\toprule
$n$ & $\chiord{2}(n,0)$ & absolute difference \\
\midrule
$3$  & $\phantom{0}{-}83.700557869890610096$  & $4.7\times10^{-38}$ \\
$9$  & $-152.72606199063291936$  & $3.1\times10^{-37}$ \\
$17$ & $-185.45063313213760688$  & $6.9\times10^{-37}$ \\
$33$ & $-217.89370199705380186$  & $6.5\times10^{-37}$ \\
$51$ & $-238.83542176998078148$  & $1.3\times10^{-33}$ \\
$71$ & $-254.67246951278298666$  & $4.1\times10^{-27}$ \\
$91$ & $-266.52751132569331901$  & $3.3\times10^{-22}$ \\
\bottomrule
\end{tabular}
\caption{$\chiord{2}(n,0)$ from the closed form against the Quantum Spectral Curve
values of Ref.~\cite{Alfimov:2018cms}. The eigenvalue column is rounded at the
last digit shown, and the reference values agree with it at every digit printed. The
difference is the residual of the recorded evaluation at forty digits of working
precision, at the settings given in \texttt{table1.json}, and it does not bound the
agreement: through $n=33$ it is set by that
working precision and at the higher spins by the cutoff of the direct sum.
\texttt{intercepts\_check.csv} of the ancillary files lists the deeper reference
residuals at the twenty spins compared at a working precision beyond forty digits, the
supplied $n=1$ block included,
and the comparison at every spin is in the ancillary files of
Ref.~\cite{companion}. The four differences through $n=33$ come from a single evaluation
and are not expected to reproduce beyond their order of magnitude.}
\label{tab:intercepts}
\end{table}

The rules were fixed at low spin from spins already computed, and not derived from the
integrand at symbolic $n$. They are exact at each spin that was generated. The
verification covers every generated odd spin up to $n=99$ and includes the holdout spin
$n=101$; a proof at arbitrary odd $n$ is lacking.

\section{Ancillary files}
The ancillary files of the arXiv record of this paper provide the rules in
machine-readable form, in nine files and a \texttt{README.txt}.
Equations~\eqref{eq:ladderform} and~\eqref{eq:kres} refer to
\texttt{chi\_nnlo\_rules.py}, which holds the \Nshapes{} atom shapes of
Eq.~\eqref{eq:atoms}, the \Nclosing{} closing rules with their rung ranges, the \Ncomposites{} composite
functions of Eq.~\eqref{eq:ladderform} written out as combinations of atoms, the rules of
the \Nomega{} nested atoms of Eq.~\eqref{eq:kres}, the closed form of Eq.~\eqref{eq:KR3} and
the regularity identity of Eq.~\eqref{eq:laurent}, all in exact rational arithmetic and
with nothing tabulated per spin. It needs the Python~3 standard library and nothing else,
and returns every coefficient of $\Fn$ at any odd spin $n\ge3$. The check at $n=1$
reported in Sec.~\ref{sec:checks} lies below that range, and none of the programs
supplied here carries it out.
\texttt{check\_compact\_form.py} expands Eqs.~\eqref{eq:ladderform}
and~\eqref{eq:kres} through that file and compares the result against the stored atom
table cell by cell in exact arithmetic, over all \Nfamilies{} families, and it performs
three further tests, listed in \texttt{README.txt}. The transcription
into \texttt{chi\_nnlo\_rules.py} is by hand. The same two displays are printed in
Ref.~\cite{companion}, and in the comparison reported there they are read out of
the manuscript source and not transcribed.

From those coefficients \texttt{chi\_nnlo\_eval.py} assembles Eq.~\eqref{eq:master}
and evaluates $\chiord{2}(n,\nu)$ in Python~3 with \texttt{mpmath}, reproducing the
eigenvalue entries of Table~\ref{tab:intercepts}. The independently extracted
coefficients for odd $n\le33$ are in \texttt{atoms\_odd\_n1\_33.json} as rationals
multiplying $1$, $\pi^{2}$ and $\zeta_3$, and those at $n=51$, $71$ and $91$, the three
spins above the range $n\le49$ the closing rules were fixed on, in
\texttt{atoms\_n51.txt}, \texttt{atoms\_n71.txt} and \texttt{atoms\_n91.txt}. The
comparison is made against these four files, the first of which also supplies the
boundary block at $n=1$. In \texttt{intercepts\_check.csv} we
list the Quantum Spectral Curve value at every odd spin through $n=91$ and the
comparison with the closed form at the twenty spins where it was carried out at a
working precision beyond forty digits, the supplied $n=1$ block included, and in
\texttt{table1.json} the rows printed
here. The formats are documented in \texttt{README.txt}, with the domain on which the
evaluator is validated and the one command that reproduces an entry of
Table~\ref{tab:intercepts}. The per-spin atom tables for every odd $n\le101$, together
with the validation record, are in the ancillary files of Ref.~\cite{companion}.

\section{Outlook}
The rules have the status stated in Sec.~\ref{sec:checks}, and it remains to derive them
from the three-loop kernel, through a radial integrand at symbolic $n$.

For one of the \Nresidual{} families of $\Kres$ Sec.~\ref{sec:coeff} excludes every
coefficient rule in harmonic sums of the two distances with spin-independent rational
coefficients, and that exclusion applies to the eigenvalue coefficient only through the
agreement of rule and coefficient, verified over the computed spins and not proved.

The ladder lengthens with the spin, and with it the number of shifted instances the
block expands into, while the \Nshapes{} shapes and the \Nfamilies{} families do not
grow. Whether the whole odd-$n$ tower
resums into a fixed-length expression in the single pair $(z,\zb)$ is left open.

Even conformal spin is untouched by the present statements, and we do not offer the
continuation of the closed form in $n$ as a route to that sector. We have reduced it
separately. That calculation has not been checked outside the
present authors and will be published elsewhere.

If the maximal-transcendentality principle, which holds at the next-to-leading
order, extends one order higher, the odd-$n$ closed form determines, for the
dimensional-reduction coupling used here, the highest-weight part of the
corresponding eigenvalue in QCD at this order.

\appendix
\numberwithin{table}{section}
\section{The coefficient catalog}
\label{app:catalog}
The rules collected here are the reconstruction of Sec.~\ref{sec:coeff}, and their
status is the one stated in Sec.~\ref{sec:checks}.
Every closing coefficient of Eq.~\eqref{eq:master} is a rational combination of
\Nweights{} functions of the two distances of Eq.~\eqref{eq:distances}, with $\epsk$
the sign the rung at $z+k$ carries. They are
\begin{align}
\wgt{1}&=1, & \wgt{2}&=\epsk, & \wgt{3}&=\epsk\,\Sh{1}{\ap},
\nonumber\\
\wgt{4}&=\epsk\,\Sh{1}{\am}, & \wgt{5}&=\epsk\,\Sh{1}{\amin}, & \wgt{6}&=\epsk\,\Sh{2}{\ap},
\nonumber\\
\wgt{7}&=\epsk\,\Sh{-2}{\ap},
\label{eq:weights}
\end{align}

and a closing rule is a fixed rational vector against them, the same vector at every
spin.

The \Nshapes{} shapes of Eq.~\eqref{eq:atoms}, \Nshapesrat{} rational and
\Nshapesnest{} nested, are listed in Table~\ref{tab:shapes} as functions of the
argument $w$. Neither the list nor its length changes with the spin.
{\small
\begin{longtable}{@{}rl@{\qquad}rl@{}}
\caption{The \Nshapes{} atom shapes of Eq.~\eqref{eq:atoms}, as functions of the argument $w$: \Nshapesrat{} rational and \Nshapesnest{} nested. The list does not grow with the spin.}\label{tab:shapes}\\
\toprule
\endfirsthead
\toprule
\endhead
\bottomrule
\endlastfoot
1 & $\frac{1}{w}$ & 22 & $\sum_{N\ge1}\frac{(-1)^{N}}{N^{3}(N+w)^{2}}$ \\
2 & $\frac{1}{w^{2}}$ & 23 & $\sum_{N\ge1}\frac{(-1)^{N}}{N^{4}(N+w)}$ \\
3 & $\frac{1}{w^{3}}$ & 24 & $\sum_{N\ge1}\frac{(-1)^{N}\,\Sh{-3}{N}}{N(N+w)}$ \\
4 & $\frac{1}{w^{4}}$ & 25 & $\sum_{N\ge1}\frac{(-1)^{N}\,\Sh{-2}{N}}{N(N+w)}$ \\
5 & $\frac{1}{w^{5}}$ & 26 & $\sum_{N\ge1}\frac{(-1)^{N}\,\Sh{-2}{N}}{N(N+w)^{2}}$ \\
6 & $\sum_{N\ge1}\frac{1}{(N+w)^{5}}$ & 27 & $\sum_{N\ge1}\frac{(-1)^{N}\,\Sh{-2,1}{N}}{N(N+w)}$ \\
7 & $\sum_{N\ge1}\frac{1}{N(N+w)}$ & 28 & $\sum_{N\ge1}\frac{(-1)^{N}\,\Sh{1}{N}}{N(N+w)^{2}}$ \\
8 & $\sum_{N\ge1}\frac{1}{N^{2}(N+w)}$ & 29 & $\sum_{N\ge1}\frac{(-1)^{N}\,\Sh{1}{N}}{N(N+w)^{3}}$ \\
9 & $\sum_{N\ge1}\frac{1}{N^{2}(N+w)^{3}}$ & 30 & $\sum_{N\ge1}\frac{(-1)^{N}\,\Sh{1}{N}}{N^{2}(N+w)}$ \\
10 & $\sum_{N\ge1}\frac{1}{N^{3}(N+w)}$ & 31 & $\sum_{N\ge1}\frac{(-1)^{N}\,\Sh{1}{N}}{N^{2}(N+w)^{2}}$ \\
11 & $\sum_{N\ge1}\frac{1}{N^{3}(N+w)^{2}}$ & 32 & $\sum_{N\ge1}\frac{(-1)^{N}\,\Sh{1}{N}}{N^{3}(N+w)}$ \\
12 & $\sum_{N\ge1}\frac{1}{N^{4}(N+w)}$ & 33 & $\sum_{N\ge1}\frac{(-1)^{N}\,\Sh{1,-2}{N}}{N(N+w)}$ \\
13 & $\sum_{N\ge1}\frac{\Sh{1}{N}}{N^{3}(N+w)}$ & 34 & $\sum_{N\ge1}\frac{(-1)^{N}\,\Sh{1,1}{N}}{N(N+w)^{2}}$ \\
14 & $\sum_{N\ge1}\frac{(-1)^{N}}{N(N+w)}$ & 35 & $\sum_{N\ge1}\frac{(-1)^{N}\,\Sh{1,1}{N}}{N^{2}(N+w)}$ \\
15 & $\sum_{N\ge1}\frac{(-1)^{N}}{N(N+w)^{2}}$ & 36 & $\sum_{N\ge1}\frac{(-1)^{N}\,\Sh{1,2}{N}}{N(N+w)}$ \\
16 & $\sum_{N\ge1}\frac{(-1)^{N}}{N(N+w)^{3}}$ & 37 & $\sum_{N\ge1}\frac{(-1)^{N}\,\Sh{2}{N}}{N(N+w)}$ \\
17 & $\sum_{N\ge1}\frac{(-1)^{N}}{N(N+w)^{4}}$ & 38 & $\sum_{N\ge1}\frac{(-1)^{N}\,\Sh{2}{N}}{N(N+w)^{2}}$ \\
18 & $\sum_{N\ge1}\frac{(-1)^{N}}{N^{2}(N+w)}$ & 39 & $\sum_{N\ge1}\frac{(-1)^{N}\,\Sh{2}{N}}{N^{2}(N+w)}$ \\
19 & $\sum_{N\ge1}\frac{(-1)^{N}}{N^{2}(N+w)^{2}}$ & 40 & $\sum_{N\ge1}\frac{(-1)^{N}\,\Sh{2,1}{N}}{N(N+w)}$ \\
20 & $\sum_{N\ge1}\frac{(-1)^{N}}{N^{2}(N+w)^{3}}$ & 41 & $\sum_{N\ge1}\frac{(-1)^{N}\,\Sh{3}{N}}{N(N+w)}$ \\
21 & $\sum_{N\ge1}\frac{(-1)^{N}}{N^{3}(N+w)}$ & & \\
\end{longtable}
}

Table~\ref{tab:rules} gives the \Nclosing{} closing rules. Each row is one shape paired
with one of $1$, $\pi^{2}$ and $\zeta_3$, and gives the rational multiple of each
ladder weight together with the rung range of the rule. With Table~\ref{tab:shapes} and
Eq.~\eqref{eq:weights} it fixes every closing coefficient of $\Fn$ at any odd $n\ge3$
without reference to a stored table. Equation~\eqref{eq:ladderform} runs its ladder
sums over the whole ladder, while four of the rules on rational shapes stop one or two
rungs short of the far end. The extra rungs contribute nothing, since $\Sh{1}{\ap}$
vanishes at $\ap=0$ and $\Sh{2}{\ap}+\Sh{-2}{\ap}$ at $\ap=0$ and $\ap=1$.
{\small
\begin{longtable}{@{}llcl@{}}
\caption{The \Nclosing{} closing coefficient rules. Each row gives the coefficient of one shape paired with one constant, as a rational combination of the weights $\wgt{1},\dots,\wgt{7}$ of Eq.~\eqref{eq:weights}. The third column gives the rung range of the rule, ``full'' meaning the whole ladder $-(n-1)\le k\le0$.}\label{tab:rules}\\
\toprule
shape & const. & rungs & rule \\
\midrule
\endfirsthead
\toprule
shape & const. & rungs & rule \\
\midrule
\endhead
\bottomrule
\endlastfoot
$\sum_{N\ge1}\frac{1}{N^{2}(N+w)}$ & $\pi^{2}$ & full & $-\tfrac{16}{3}\,\wgt{2}$ \\
$\sum_{N\ge1}\frac{1}{N^{2}(N+w)^{3}}$ & $1$ & full & $-32\,\wgt{2}$ \\
$\sum_{N\ge1}\frac{1}{N^{3}(N+w)}$ & $1$ & full & $-32\,\wgt{3} +96\,\wgt{4} -32\,\wgt{5}$ \\
$\sum_{N\ge1}\frac{1}{N^{3}(N+w)^{2}}$ & $1$ & full & $-48\,\wgt{2}$ \\
$\sum_{N\ge1}\frac{1}{N^{4}(N+w)}$ & $1$ & full & $-96\,\wgt{2}$ \\
$\sum_{N\ge1}\frac{1}{(N+w)^{5}}$ & $1$ & $k=1$ & $32\,\wgt{1}$ \\
$\sum_{N\ge1}\frac{\Sh{1}{N}}{N^{3}(N+w)}$ & $1$ & full & $-64\,\wgt{2}$ \\
$\sum_{N\ge1}\frac{(-1)^{N}}{N(N+w)}$ & $\zeta_3$ & full & $-32\,\wgt{2}$ \\
$\sum_{N\ge1}\frac{(-1)^{N}}{N(N+w)^{2}}$ & $\pi^{2}$ & full & $-\tfrac{16}{3}\,\wgt{2}$ \\
$\sum_{N\ge1}\frac{(-1)^{N}}{N(N+w)^{3}}$ & $1$ & full & $-32\,\wgt{3} -32\,\wgt{4} +32\,\wgt{5}$ \\
$\sum_{N\ge1}\frac{(-1)^{N}}{N(N+w)^{4}}$ & $1$ & full & $-48\,\wgt{2}$ \\
$\sum_{N\ge1}\frac{(-1)^{N}}{N^{2}(N+w)}$ & $\pi^{2}$ & full & $-\tfrac{16}{3}\,\wgt{2}$ \\
$\sum_{N\ge1}\frac{(-1)^{N}}{N^{2}(N+w)^{2}}$ & $1$ & full & $-96\,\wgt{3} -32\,\wgt{4} +64\,\wgt{5}$ \\
$\sum_{N\ge1}\frac{(-1)^{N}}{N^{2}(N+w)^{3}}$ & $1$ & full & $-64\,\wgt{2}$ \\
$\sum_{N\ge1}\frac{(-1)^{N}}{N^{3}(N+w)}$ & $1$ & full & $-160\,\wgt{3} -32\,\wgt{4} +96\,\wgt{5}$ \\
$\sum_{N\ge1}\frac{(-1)^{N}}{N^{3}(N+w)^{2}}$ & $1$ & full & $-144\,\wgt{2}$ \\
$\sum_{N\ge1}\frac{(-1)^{N}}{N^{4}(N+w)}$ & $1$ & full & $-288\,\wgt{2}$ \\
$\sum_{N\ge1}\frac{(-1)^{N}\,\Sh{-2,1}{N}}{N(N+w)}$ & $1$ & full & $128\,\wgt{2}$ \\
$\sum_{N\ge1}\frac{(-1)^{N}\,\Sh{-2}{N}}{N(N+w)}$ & $1$ & full & $-64\,\wgt{3} -64\,\wgt{4} +64\,\wgt{5}$ \\
$\sum_{N\ge1}\frac{(-1)^{N}\,\Sh{-2}{N}}{N(N+w)^{2}}$ & $1$ & full & $-32\,\wgt{2}$ \\
$\sum_{N\ge1}\frac{(-1)^{N}\,\Sh{-3}{N}}{N(N+w)}$ & $1$ & full & $-96\,\wgt{2}$ \\
$\sum_{N\ge1}\frac{(-1)^{N}\,\Sh{1,-2}{N}}{N(N+w)}$ & $1$ & full & $-64\,\wgt{2}$ \\
$\sum_{N\ge1}\frac{(-1)^{N}\,\Sh{1,1}{N}}{N(N+w)^{2}}$ & $1$ & full & $-128\,\wgt{2}$ \\
$\sum_{N\ge1}\frac{(-1)^{N}\,\Sh{1,1}{N}}{N^{2}(N+w)}$ & $1$ & full & $-128\,\wgt{2}$ \\
$\sum_{N\ge1}\frac{(-1)^{N}\,\Sh{1,2}{N}}{N(N+w)}$ & $1$ & full & $-128\,\wgt{2}$ \\
$\sum_{N\ge1}\frac{(-1)^{N}\,\Sh{1}{N}}{N(N+w)^{2}}$ & $1$ & full & $64\,\wgt{3} +64\,\wgt{4} -64\,\wgt{5}$ \\
$\sum_{N\ge1}\frac{(-1)^{N}\,\Sh{1}{N}}{N(N+w)^{3}}$ & $1$ & full & $32\,\wgt{2}$ \\
$\sum_{N\ge1}\frac{(-1)^{N}\,\Sh{1}{N}}{N^{2}(N+w)}$ & $1$ & full & $64\,\wgt{3} +64\,\wgt{4} -64\,\wgt{5}$ \\
$\sum_{N\ge1}\frac{(-1)^{N}\,\Sh{1}{N}}{N^{2}(N+w)^{2}}$ & $1$ & full & $96\,\wgt{2}$ \\
$\sum_{N\ge1}\frac{(-1)^{N}\,\Sh{1}{N}}{N^{3}(N+w)}$ & $1$ & full & $160\,\wgt{2}$ \\
$\sum_{N\ge1}\frac{(-1)^{N}\,\Sh{2,1}{N}}{N(N+w)}$ & $1$ & full & $-128\,\wgt{2}$ \\
$\sum_{N\ge1}\frac{(-1)^{N}\,\Sh{2}{N}}{N(N+w)}$ & $1$ & full & $64\,\wgt{3} +64\,\wgt{4} -64\,\wgt{5}$ \\
$\sum_{N\ge1}\frac{(-1)^{N}\,\Sh{2}{N}}{N(N+w)^{2}}$ & $1$ & full & $96\,\wgt{2}$ \\
$\sum_{N\ge1}\frac{(-1)^{N}\,\Sh{2}{N}}{N^{2}(N+w)}$ & $1$ & full & $160\,\wgt{2}$ \\
$\sum_{N\ge1}\frac{(-1)^{N}\,\Sh{3}{N}}{N(N+w)}$ & $1$ & full & $160\,\wgt{2}$ \\
$\frac{1}{w}$ & $\pi^{2}$ & $k\ge-(n-3)$ & $\tfrac{16}{3}\,\wgt{6} +\tfrac{16}{3}\,\wgt{7}$ \\
$\frac{1}{w}$ & $\zeta_3$ & $k\ge-(n-2)$ & $32\,\wgt{3}$ \\
$\frac{1}{w^{2}}$ & $\pi^{2}$ & $k\ge-(n-2)$ & $\tfrac{16}{3}\,\wgt{3}$ \\
$\frac{1}{w^{4}}$ & $1$ & $k\ge-(n-2)$ & $48\,\wgt{3}$ \\
$\frac{1}{w^{5}}$ & $1$ & $k=1$ & $32\,\wgt{1}$ \\
\end{longtable}
}

The \Ncomposites{} functions of one variable that Eq.~\eqref{eq:ladderform} collects
\Ncomposed{} of the \Nclosing{} rules into are combinations of the atoms at a single
argument,
\begin{align}
\Gladder_{(-1)^k}(w)&=
  -32\,\sum_{N\ge1}\frac{1}{N^{2}(N+w)^{3}} -48\,\sum_{N\ge1}\frac{1}{N^{3}(N+w)^{2}}
  \nonumber\\ &\quad -96\,\sum_{N\ge1}\frac{1}{N^{4}(N+w)} -64\,\sum_{N\ge1}\frac{\Sh{1}{N}}{N^{3}(N+w)}
  \nonumber\\ &\quad -48\,\sum_{N\ge1}\frac{(-1)^{N}}{N(N+w)^{4}} -64\,\sum_{N\ge1}\frac{(-1)^{N}}{N^{2}(N+w)^{3}}
  \nonumber\\ &\quad -144\,\sum_{N\ge1}\frac{(-1)^{N}}{N^{3}(N+w)^{2}} -288\,\sum_{N\ge1}\frac{(-1)^{N}}{N^{4}(N+w)}
  \nonumber\\ &\quad +128\,\sum_{N\ge1}\frac{(-1)^{N}\,\Sh{-2,1}{N}}{N(N+w)} -32\,\sum_{N\ge1}\frac{(-1)^{N}\,\Sh{-2}{N}}{N(N+w)^{2}}
  \nonumber\\ &\quad -96\,\sum_{N\ge1}\frac{(-1)^{N}\,\Sh{-3}{N}}{N(N+w)} -64\,\sum_{N\ge1}\frac{(-1)^{N}\,\Sh{1,-2}{N}}{N(N+w)}
  \nonumber\\ &\quad -128\,\sum_{N\ge1}\frac{(-1)^{N}\,\Sh{1,1}{N}}{N(N+w)^{2}} -128\,\sum_{N\ge1}\frac{(-1)^{N}\,\Sh{1,1}{N}}{N^{2}(N+w)}
  \nonumber\\ &\quad -128\,\sum_{N\ge1}\frac{(-1)^{N}\,\Sh{1,2}{N}}{N(N+w)} +32\,\sum_{N\ge1}\frac{(-1)^{N}\,\Sh{1}{N}}{N(N+w)^{3}}
  \nonumber\\ &\quad +96\,\sum_{N\ge1}\frac{(-1)^{N}\,\Sh{1}{N}}{N^{2}(N+w)^{2}} +160\,\sum_{N\ge1}\frac{(-1)^{N}\,\Sh{1}{N}}{N^{3}(N+w)}
  \nonumber\\ &\quad -128\,\sum_{N\ge1}\frac{(-1)^{N}\,\Sh{2,1}{N}}{N(N+w)} +96\,\sum_{N\ge1}\frac{(-1)^{N}\,\Sh{2}{N}}{N(N+w)^{2}}
  \nonumber\\ &\quad +160\,\sum_{N\ge1}\frac{(-1)^{N}\,\Sh{2}{N}}{N^{2}(N+w)} +160\,\sum_{N\ge1}\frac{(-1)^{N}\,\Sh{3}{N}}{N(N+w)}
  \nonumber\\ &\quad -\tfrac{16}{3}\,\pi^{2}\,\sum_{N\ge1}\frac{1}{N^{2}(N+w)} -\tfrac{16}{3}\,\pi^{2}\,\sum_{N\ge1}\frac{(-1)^{N}}{N(N+w)^{2}}
  \nonumber\\ &\quad -\tfrac{16}{3}\,\pi^{2}\,\sum_{N\ge1}\frac{(-1)^{N}}{N^{2}(N+w)} -32\,\zeta_3\,\sum_{N\ge1}\frac{(-1)^{N}}{N(N+w)}
\nonumber\\[6pt]
\Phi(w)&=
  -64\,\sum_{N\ge1}\frac{(-1)^{N}\,\Sh{-2}{N}}{N(N+w)} +64\,\sum_{N\ge1}\frac{(-1)^{N}\,\Sh{1}{N}}{N(N+w)^{2}}
  \nonumber\\ &\quad +64\,\sum_{N\ge1}\frac{(-1)^{N}\,\Sh{1}{N}}{N^{2}(N+w)} +64\,\sum_{N\ge1}\frac{(-1)^{N}\,\Sh{2}{N}}{N(N+w)}
\nonumber\\[6pt]
D_{+}(w)&=
  -32\,\sum_{N\ge1}\frac{1}{N^{3}(N+w)} -32\,\sum_{N\ge1}\frac{(-1)^{N}}{N(N+w)^{3}}
  \nonumber\\ &\quad -96\,\sum_{N\ge1}\frac{(-1)^{N}}{N^{2}(N+w)^{2}} -160\,\sum_{N\ge1}\frac{(-1)^{N}}{N^{3}(N+w)}
  \nonumber\\ &\quad +48\,\frac{1}{w^{4}} +\tfrac{16}{3}\,\pi^{2}\,\frac{1}{w^{2}}
  \nonumber\\ &\quad +32\,\zeta_3\,\frac{1}{w}
\nonumber\\[6pt]
D_{-}(w)&=
  96\,\sum_{N\ge1}\frac{1}{N^{3}(N+w)} -32\,\sum_{N\ge1}\frac{(-1)^{N}}{N(N+w)^{3}}
  \nonumber\\ &\quad -32\,\sum_{N\ge1}\frac{(-1)^{N}}{N^{2}(N+w)^{2}} -32\,\sum_{N\ge1}\frac{(-1)^{N}}{N^{3}(N+w)}
\nonumber\\[6pt]
D_{\min}(w)&=
  -32\,\sum_{N\ge1}\frac{1}{N^{3}(N+w)} +32\,\sum_{N\ge1}\frac{(-1)^{N}}{N(N+w)^{3}}
  \nonumber\\ &\quad +64\,\sum_{N\ge1}\frac{(-1)^{N}}{N^{2}(N+w)^{2}} +96\,\sum_{N\ge1}\frac{(-1)^{N}}{N^{3}(N+w)}
\label{eq:composites}
\end{align}

which is \Ncompterms{} terms in all.

Of the \Nresidual{} rules of $\Kres$, Eqs.~\eqref{eq:cd} and~\eqref{eq:cC} give those on
the alternating pair and on $\Om{C}$, and Eqs.~\eqref{eq:KR3} and~\eqref{eq:laurent}
return the three rational ones. The remaining one is the exchange-odd rule, which takes
weight three. With
\begin{equation}
\Lin{j}(\ap,\am)=\sum_{N=1}^{\ap}\frac{\Sh{1}{\min(\am+N,\ap-N)}}{N^{j}},\qquad
\Lnul{j}(\ap)=\sum_{N=1}^{\ap}\frac{\Sh{1}{\ap-N}}{N^{j}},
\label{eq:laddersummin}
\end{equation}
the sums of Eq.~\eqref{eq:laddersum} with $\min$ for $\max$, the second of them the
first at $\am=\ap$, where $\min(\am+N,\ap-N)=\ap-N$ at every $N$, it reads
\begin{align}
\cOm(n,k)&=\epsk\bigl[\varphi(\ap,\am)-\varphi(\am,\ap)\bigr],
\nonumber\\[4pt]
\varphi(x,y)&=192\,\Lnul{2}(x)-96\,\Lin{2}(x,y)-128\,\Lax{2}(x,y)
\nonumber\\
&\quad-64\,\Sh{1,-2}{x}-32\,\Sh{1}{x}\Sh{2}{x}+128\,\Sh{3}{x}+32\,\Sh{-3}{x}.
\label{eq:cO}
\end{align}
The antisymmetry in the two distances is manifest, and $\varphi$ carries the weight
three the rule needs.

Tables~\ref{tab:shapes} and~\ref{tab:rules} with Eqs.~\eqref{eq:weights},
\eqref{eq:composites}, \eqref{eq:laddersum}, \eqref{eq:laddersummin},
\eqref{eq:cd}, \eqref{eq:cC}, \eqref{eq:cO}, \eqref{eq:KR3}
and~\eqref{eq:laurent} are the whole specification of the coefficients that enter the
main result of Sec.~\ref{sec:coeff}, all \Nfamilies{} families at every
odd $n\ge3$, with nothing tabulated per spin. They are the same rules
\texttt{chi\_nnlo\_rules.py} holds, and the tables here are generated from that file.

\section*{Acknowledgments}
We thank Sergey Bondarenko for valuable discussions. The symbolic reduction and the
numerical evaluations were carried out by the authors with computer algebra and with
programs written for this purpose. All scientific content, derivations
and conclusions were developed and verified by the authors.

During the preparation of this work the authors used Opus 4.8, Opus 5, Fable 5 and
Fable 5.1 (Anthropic), GPT-5.5, GPT-5.6-sol and GPT-6-astra (OpenAI), and Gemini 3.1 Pro
(Google) in preparing the accompanying programs and in language editing. The authors
reviewed and edited all of it and take full responsibility for the content.

\section*{CRediT authorship contribution statement}
A. Prygarin: conceptualization, methodology, software, validation, investigation,
writing (original draft), writing (review and editing).
C. C. Madjuogang Sandeu: conceptualization, methodology, software, validation,
investigation, writing (original draft), writing (review and editing).

\section*{Funding}
This work is supported in part by the ``Program of HEP support---Council of Higher
Education of Israel.''

\section*{Declaration of competing interest}
The authors declare that they have no known competing financial interests or personal
relationships that could have appeared to influence the work reported in this paper.



\begin{thebibliography}{99}


\bibitem{Fadin:1975cb}
V.~S.~Fadin, E.~A.~Kuraev and L.~N.~Lipatov,
``On the Pomeranchuk Singularity in Asymptotically Free Theories,''
Phys.\ Lett.\ B \textbf{60} (1975) 50.

\bibitem{Kuraev:1977fs}
E.~A.~Kuraev, L.~N.~Lipatov and V.~S.~Fadin,
``The Pomeranchuk singularity in nonabelian gauge theories,''
Sov.\ Phys.\ JETP \textbf{45} (1977) 199.

\bibitem{Balitsky:1978ic}
I.~I.~Balitsky and L.~N.~Lipatov,
``The Pomeranchuk Singularity in Quantum Chromodynamics,''
Sov.\ J.\ Nucl.\ Phys.\ \textbf{28} (1978) 822.

\bibitem{Lipatov:1985uk}
L.~N.~Lipatov,
``The Bare Pomeron in Quantum Chromodynamics,''
Sov.\ Phys.\ JETP \textbf{63} (1986) 904.

\bibitem{Lipatov:1993yb}
L.~N.~Lipatov,
``Asymptotic behavior of multicolor QCD at high energies in connection with exactly solvable spin models,''
JETP Lett.\ \textbf{59} (1994) 596
[arXiv:hep-th/9311037].

\bibitem{Faddeev:1994zg}
L.~D.~Faddeev and G.~P.~Korchemsky,
``High energy QCD as a completely integrable model,''
Phys.\ Lett.\ B \textbf{342} (1995) 311
[arXiv:hep-th/9404173].

\bibitem{Kotikov:2002ab}
A.~V.~Kotikov and L.~N.~Lipatov,
``DGLAP and BFKL equations in the $N=4$ supersymmetric gauge theory,''
Nucl.\ Phys.\ B \textbf{661} (2003) 19
[Erratum: Nucl.\ Phys.\ B \textbf{685} (2004) 405]
[arXiv:hep-ph/0208220].

\bibitem{Costa:2012cb}
M.~S.~Costa, V.~Gon\c{c}alves and J.~Penedones,
``Conformal Regge theory,''
JHEP \textbf{12} (2012) 091
[arXiv:1209.4355].

\bibitem{Alfimov:2018cms}
M.~Alfimov, N.~Gromov and G.~Sizov,
``BFKL spectrum of $N=4$: non-zero conformal spin,''
JHEP \textbf{07} (2018) 181
[arXiv:1802.06908].

\bibitem{Fadin:1998py}
V.~S.~Fadin and L.~N.~Lipatov,
``BFKL pomeron in the next-to-leading approximation,''
Phys.\ Lett.\ B \textbf{429} (1998) 127
[arXiv:hep-ph/9802290].

\bibitem{Ciafaloni:1998gs}
M.~Ciafaloni and G.~Camici,
``Energy scale(s) and next-to-leading BFKL equation,''
Phys.\ Lett.\ B \textbf{430} (1998) 349
[arXiv:hep-ph/9803389].

\bibitem{Kotikov:2000pm}
A.~V.~Kotikov and L.~N.~Lipatov,
``NLO corrections to the BFKL equation in QCD and in supersymmetric gauge theories,''
Nucl.\ Phys.\ B \textbf{582} (2000) 19
[arXiv:hep-ph/0004008].

\bibitem{Kotikov:2006ts}
A.~V.~Kotikov and L.~N.~Lipatov,
``On the highest transcendentality in $N=4$ SUSY,''
Nucl.\ Phys.\ B \textbf{769} (2007) 217
[arXiv:hep-th/0611204].

\bibitem{Bondarenko:2015tba}
S.~Bondarenko and A.~Prygarin,
``Hermitian separability and transition from singlet to adjoint BFKL equations in $N=4$ super Yang--Mills Theory,''
[arXiv:1510.00589].

\bibitem{Joubat:2020hvc}
M.~Joubat and A.~Prygarin,
``Hermitian separability of BFKL eigenvalue in Bethe--Salpeter approach,''
Eur.\ Phys.\ J.\ C \textbf{80} (2020) 1183
[arXiv:2007.15388].

\bibitem{Caron-Huot:2016tzz}
S.~Caron-Huot and M.~Herranen,
``High-energy evolution to three loops,''
JHEP \textbf{02} (2018) 058
[arXiv:1604.07417].

\bibitem{Brunello:2025rhh}
G.~Brunello, S.~Caron-Huot, G.~Crisanti, M.~Giroux and S.~Smith,
``High-energy evolution in planar QCD to three loops: the non-conformal contribution,''
JHEP \textbf{11} (2025) 055
[arXiv:2508.03794].

\bibitem{Gromov:2013pga}
N.~Gromov, V.~Kazakov, S.~Leurent and D.~Volin,
``Quantum Spectral Curve for Planar $N=4$ Super-Yang--Mills Theory,''
Phys.\ Rev.\ Lett.\ \textbf{112} (2014) 011602
[arXiv:1305.1939].

\bibitem{Alfimov:2014bwa}
M.~Alfimov, N.~Gromov and V.~Kazakov,
``QCD Pomeron from AdS/\allowbreak CFT Quantum Spectral Curve,''
JHEP \textbf{07} (2015) 164
[arXiv:1408.2530].

\bibitem{Gromov:2015vua}
N.~Gromov, F.~Levkovich-Maslyuk and G.~Sizov,
``Pomeron Eigenvalue at Three Loops in $N=4$ Supersymmetric Yang--Mills Theory,''
Phys.\ Rev.\ Lett.\ \textbf{115} (2015) 251601
[arXiv:1507.04010].

\bibitem{Velizhanin:2015xsa}
V.~N.~Velizhanin,
``BFKL pomeron in the next-to-next-to-leading approximation in the planar
$N=4$ SYM theory,''
[arXiv:1508.02857].

\bibitem{Velizhanin:2021bdh}
V.~N.~Velizhanin,
``NNNLLA BFKL pomeron eigenvalue in the planar $N=4$ SYM theory,''
[arXiv:2106.06527].

\bibitem{Ekhammar:2024neh}
S.~Ekhammar, N.~Gromov and M.~Preti,
``Long Range Asymptotic Baxter-Bethe Ansatz for $N=4$ BFKL,''
[arXiv:2406.18639].

\bibitem{Ekhammar:2025vig}
S.~Ekhammar, N.~Gromov and M.~Preti,
``Regge trajectories of $N=4$ SYM. Part I. General Asymptotic Baxter-Bethe Ansatz,''
JHEP \textbf{02} (2026) 027
[arXiv:2507.15983].

\bibitem{Joubat:2020vrw}
M.~Joubat and A.~Prygarin,
``Reflection Identities of Harmonic Sums and pole decomposition of BFKL eigenvalue,''
Int.\ J.\ Mod.\ Phys.\ A \textbf{36} (2021) 2150025
[arXiv:2011.08095].

\bibitem{Dixon:2012yy}
L.~J.~Dixon, C.~Duhr and J.~Pennington,
``Single-valued harmonic polylogarithms and the multi-Regge limit,''
JHEP \textbf{10} (2012) 074
[arXiv:1207.0186].

\bibitem{Dixon:2014voa}
L.~J.~Dixon, J.~M.~Drummond, C.~Duhr and J.~Pennington,
``The four-loop remainder function and multi-Regge behavior at NNLLA in
planar $N=4$ super-Yang--Mills theory,''
JHEP \textbf{06} (2014) 116
[arXiv:1402.3300].

\bibitem{Basso:2014pla}
B.~Basso, S.~Caron-Huot and A.~Sever,
``Adjoint BFKL at finite coupling: a short-cut from the collinear limit,''
JHEP \textbf{01} (2015) 027
[arXiv:1407.3766].

\bibitem{Fadin:2011we}
V.~S.~Fadin and L.~N.~Lipatov,
``BFKL equation for the adjoint representation of the gauge group in the
next-to-leading approximation at $N=4$ SUSY,''
Phys.\ Lett.\ B \textbf{706} (2012) 470
[arXiv:1111.0782].

\bibitem{Joubat:2021nss}
M.~Joubat and A.~Prygarin,
``Universal transcendentality limit of BFKL eigenvalue,''
Eur.\ Phys.\ J.\ C \textbf{82} (2022) 45
[arXiv:2111.11664].

\bibitem{Prygarin:2018tng}
A.~Prygarin,
``Reflection identities of harmonic sums up to weight three,''
[arXiv:1808.09307].

\bibitem{Prygarin:2018cog}
A.~Prygarin,
``Reflection identities of harmonic sums of weight four,''
Universe \textbf{5} (2019) 77
[arXiv:1809.06696].

\bibitem{Joubat:2019esj}
M.~Joubat and A.~Prygarin,
``The analytic structure of the BFKL equation and reflection identities of harmonic sums at weight five,''
Int.\ J.\ Mod.\ Phys.\ A \textbf{34} (2019) 1950064
[arXiv:1903.06773].

\bibitem{Joubat:2023ftd}
M.~Joubat, C.~C.~Madjuogang Sandeu and A.~Prygarin,
``Pole decomposition of BFKL eigenvalue at zero conformal spin and the real part of digamma function,''
Phys.\ Lett.\ B \textbf{848} (2024) 138319
[arXiv:2307.12112].

\bibitem{companion}
A.~Prygarin and C.~C.~Madjuogang Sandeu,
``The next-to-next-to-leading order BFKL eigenvalue at odd conformal spin in
planar $\mathcal{N}=4$ super Yang--Mills: a reconstructed closed form, coefficient
structure, and arithmetic,''
[arXiv:2607.17613].

\bibitem{Vermaseren:1998uu}
J.~A.~M.~Vermaseren,
``Harmonic sums, Mellin transforms and integrals,''
Int.\ J.\ Mod.\ Phys.\ A \textbf{14} (1999) 2037
[arXiv:hep-ph/9806280].

\bibitem{Salam:1998tj}
G.~P.~Salam,
``A resummation of large sub-leading corrections at small $x$,''
JHEP \textbf{07} (1998) 019
[arXiv:hep-ph/9806482].

\bibitem{Marzani:2007gk}
S.~Marzani, R.~D.~Ball, P.~Falgari and S.~Forte,
``BFKL at next-to-next-to-leading order,''
Nucl.\ Phys.\ B \textbf{783} (2007) 143
[arXiv:0704.2404].

\end{thebibliography}
\end{document}